# Controlling and securing a Digital Home using Multiple Sensor Based Perception system Integrated with Mobile and Voice technology

[1] Avishek Ahmed, [2] Tanvir Ahmed, [3] Md. Samawat Ullah, [4] Md. Manirul Islam

[1,2,3,4] Department of Computer Science and Engineering
American International University-Bangladesh, Dhaka, Bangladesh.
[1] avishek@aiub.edu, [2] tanvir@aiub.edu, [3] samawat@gmail.com, [4] manirul@aiub.edu

## ABSTRACT

Fully controlled digital home had always been considered as a luxury of rich people because of excessive cost to install the system. It is now within the reach of mass people with lots of inexpensive cool features. In this paper we have designed and developed a very low cost, efficient and reliable Digital home system. Fully Controlled Digital Home is no more a Luxury. Our proposed system made it affordable. We built a low-cost feature-rich Digital Home System (DHS). Digital Home System is combination of automated services i.e. Electronic Device Controller, IR Security System, Web Desktop, Remote Video Surveillance System and Virtual Mobile by which we can control our home by avoiding old manual processes e.g. our physical presence at home is optional. The System provides some of the modern luxury & security features to us. Now we can control Light or fan or AC or any electronic devices in home by voice command, Blue-tooth, GPRS or Website. To control the system remotely, GPRS connectivity is added. We can also monitor our home from remote area by using Remote Video Surveillance System. For example if we are outside of home but we want to see what is happening in your home you can easily login to system and watch the live video of your home in our mobile device. Moreover, we can also access our PC and do the necessary tasks from any internet enabled computer in the world by using Web Desktop which is specially built for this purpose. Furthermore, if any unauthorized person wants to enter in our home the IR-Security System will inform us by sending SMS & store the image of that unauthorized person in our computer for further action and also generate a voice alarm "Someone in the room". So that ensures the security of our valuable things. Also we can identify and monitor the location of our valuable assets e.g. precious metals remotely. Finally, Virtual mobile application is a Universal mobile Driver by which we can exactly perform some same task e.g. Remote call, Phone book access, SMS read-write of our mobile device from our new invented computer's virtual mobile. Water tank controller, automated entrance/exit gate, remote TV connection is also added in our digital home system.

**Keywords-** Digital Home System; DHS; Web desktop; IR-Security; Virtual Mobile; Remote Surveillance; Remote video Surveillance; Electronic Device Controller; Device Communicator; Interactive Home; Smart Home; Digital Home Introduction.

## 1. INTRODUCTION

Home is the most important platform in one's life –and that platform is about to change. New technology is bringing in new possibilities and changing the way people live their lives.

Human being likes luxury. There are some kinds of people who wish to live more relaxed life. For example a person returns from office & s/he wants a coolest room when s/he will reach at home or during midnight, one may not feel like get up and switch off the AC or Fan. They always have wished to get a system that would automatically handle these electronic devices for them. But that's when they feel the lack of a system like our DHS. While leaving for office or else, sometimes a person may forget to switch off the Light/fan/AC or any electronic device that cost the user for no reason. There may be a case like the person is outside of the home & s/he wants to access his/her PC or one may forget some data in his/her PC and s/he needs to access that immediately but s/he cannot do that for a system. To get the data s/he must have to come back to home. In case of Bangladeshi scenario there are so many cases where the person is in office & s/he feels insecure about his/her home that what is going on there or one may want to get the immediate response if his/her locker is accessed by any unauthorized person. There may be so many unexpected occurrences by which family member(s) may get tensed but they cannot know what is happening in their home. The person(s) feel the lack of a system like our proposed system by which s/he can see the current condition of his/her home. Some people may feel SMS typing or accessing phone book in computer is more comfortable than small mobile device. Some may forget his or her mobile at another room & wants to use at midnight, which could be solved by a system like "virtual mobile". All these issues are very important to take into account to develop the system.

In section 2 of this paper, an overview of the current trend of the technology is presented. Our proposal of a low cost digital home that uses mobile and voice technology is explained in section 3. Section 4 illustrated the simulated output of your system. The benefits of our model are explained in Section 5. Finally the paper concludes in section 6.

## 2. LITERATURE REVIEW

Various digital home system solutions are available. Most of them are very expensive. We have studied different individual components separately and integrated them together with some more innovative features to make a digital home.

Digital homes are expected to be the standard in the near future when all aspects of the home can be monitored and controlled remotely by the home owner. *"Controlling and Securing a Digital Home using Multiple Sensor Based Perception System Integrated with Mobile and Voice Technology"* research are not uncommon and a few successful example would be the *"Digital Home:* An All-in-one Device to Control most everything" [1], "eyeOS" [2], "Videosurveillance and Recording[3]", "Kaseya Remote Desktop Management[4]" from the different developer or software development firms.

Our System is different from most of the techniques described in the literature because a home owner can control and view the status of her home either on a PC, or a handheld device with an overall safety. IntuiSec [7] is a framework for intuitive user interaction with smart home security using mobile devices. The design approach of IntuiSec is to introduce a level of indirection between the user-level intent and the system-level security infrastructure. An integrated dual-level vision based home security system, establishing trusted relationships between devices, and granting temporal, selective access for both home occupants and visitors to devices within the smart home and is presented in [8]. Dual-level vision consists of two subsystems; a face recognition module and a motion detection module. The primary face recognition module works as a user authentication device; the secondary motion detection module scans for human-related movements within certain enclosed areas inside the home while the occupants are away. A Wireless USB based home security system using a gas sensor, a gas valve controller, a magnetic sensor, an infrared sensor and a digital door lock. These devices need four Wireless USB home controllers to be managed and controlled [9]. A home security system is proposed that can be built by individuals [10]. The proposed system is constructed by a general-purpose computer, cell phone and the Internet. The general-purpose PC is used as the server with some sensors and devices. The web browser on the cellular phone is used as the client. This system permits capture of an intruder's image and sends e-mails warning the homeowner of any security breaches. A new low cost multipurpose home security system, called Secure-Way, has been developed by KBM Enterprises, Inc. in Huntsville, Alabama [11]. Secure-Way models range from an effective remote control motion detector that sense the presence of an intruder and triggers an alarm, to an integrated motion system that is armed with an alarm as well as a remote video camera.

In paper [5], they proposed a technique for visualization of home safety called SaViT which stands for Safety Visualization Technique that incorporates the following three features:

1. Visualization of home safety on the user's device,
2. Justification for home safety visualization, and
3. Periodic updating of home safety visualization.

By visualizing home safety on the user's desktop, laptop, PDA or cell phone, the user is always provided with a summary view of home safety along with reasons. Moreover, appropriate implementation of this technique ensures not only platform independence but also low bandwidth consumption. SaViT technique employs the Non-Functional Requirements (NFR) Approach [6] to make summary judgment of home safety and to justify the current safety level.

The systems that we have discussed above are related to our proposed system in some ways. However there are some facts that motivated us to develop this system.

- Some works has been done for controlling Electronic devices either using Voice Command or Bluetooth or via internet. But these all features were not provided in one system. We come to a decision that we will combine these entire four features to control the electronic devices.
- There are so many products in the market on Security system (i.e. "Disaster Management System Using Wireless Technology and/or Internet" etc). But this single product is so costly. We thought about cost minimization of this product.
- We decided to develop some applications by which user can control the system and monitor the home condition of home from anywhere in the world.
- Finally we have combined all the application in one system by which user can control all the electronic devices of his/her home, ensure the security, Monitor the house 24/7 and control the system from anywhere in the world.
- We will develop all the different application and integrate all (i.e. IR security, Web desktop etc.) in one system. All the application together will make a perfect Digital Dream Home system.
- The products that are available in the market are so costly. As we combined all the system together that minimized the total cost of system development.
- Talking to appliance in a home has been a science fiction staple for decades.
- Integrate all the components home which will make a complete digital dream home
  Making a home that is cost effective. The price should be minimum but all the component will make a complete digital dream home.

## 3. PROPOSED DESIGN

*A. Architecture*

In our architecture we have some major goals and for each goal we have some specific objectives.
**GOAL 1**
Our first goal was to provide an automated system in terms of manual system to make life simpler & add some luxury features.

**Objective 1.1:** The Digital Home System will help the user to lead a luxurious life. User can control Light or fan AC or any electronic material in home by voice command, Bluetooth, GPRS, and SMS. The system will also reduce the time consumption.

**Objective 1.2:** The system will give the portability to its users. User will be able to access PC being outside of the house.

**Objective 1.3:** System will do most of the regular tasks of human being (e.g. light/fan on/off). So there will be more time to relax. Time is most important factor. This system will reduce the time consumption.

**Objective 1.4:** Unlike the manual system users does not have to present physically to that particular place. All he/she needs to do is just use his mobile phone to get the job done.

**GOAL 2**

Security is a big issue in our daily life. The system will provide some security features to its users.

**Objective 2.1:** If any unauthorized person enters, the system will inform to the Home Owner by sending **SMS** & store the image in his/her PC. The user will be able to take action immediately. Like that way security of home will increase.

**Objective 2.2:** There will be a web cam in the entrance gate of Digital Home which ensures the security. Home owner shell be able to monitor his/her house any time.

**Objective 2.3:** All the electronic devices will be controlled by the system. So the risk of electric shock shell totally reduced. People can now live a tension free life.

The use-case diagram of our proposed system is given in figure 1.

Basic Units:
- Electronics device controller
  - Electronic device controller using voice command
  - Electronic device controller using Bluetooth
  - Electronic device control using Website
- IR Security system
- Remote Video Surveillance System / Remote eye
- Web Desktop
- Virtual Mobile

The following part will describe each of the units of our proposed system.

*1) Electronics Device Controller*

The Major complexity in this subsystem is to merge computer, mobile & microcontroller and synchronization of these three appliances. To send & receive data from microcontroller we had to do it by using serial cable. So we used comm API. The system is absolutely customizable & we developed it by using object oriented approach. If any subsystem is changed, it will not affect on any component of DHS. The system components (hardware & software) are abstract from each other. So the system can be controlled by using Bluetooth or voice or website without changing hardware. The Electronic device controller software is divided into four main parts. These parts have their individual responsibilities which are listed below:

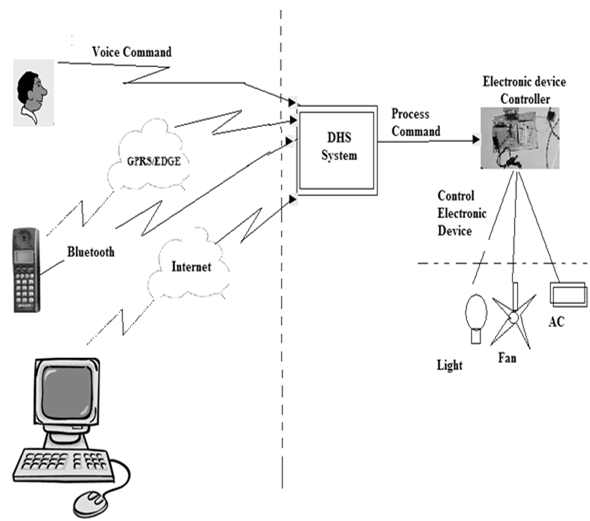

Figure 2. System Overview of electronic device controller

The activity diagram of our electronic device controller is given in figure 3.

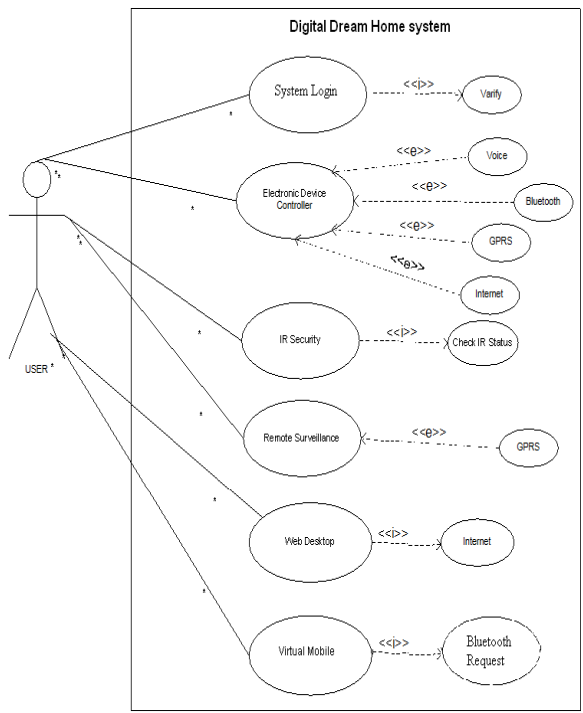

Figure 1. Use-case diagram of our system

**B. Unites**

Our proposed model contained 5 basic units. Some units contain subunits. Combining all of the units a fully controlled digital home can be achieved.

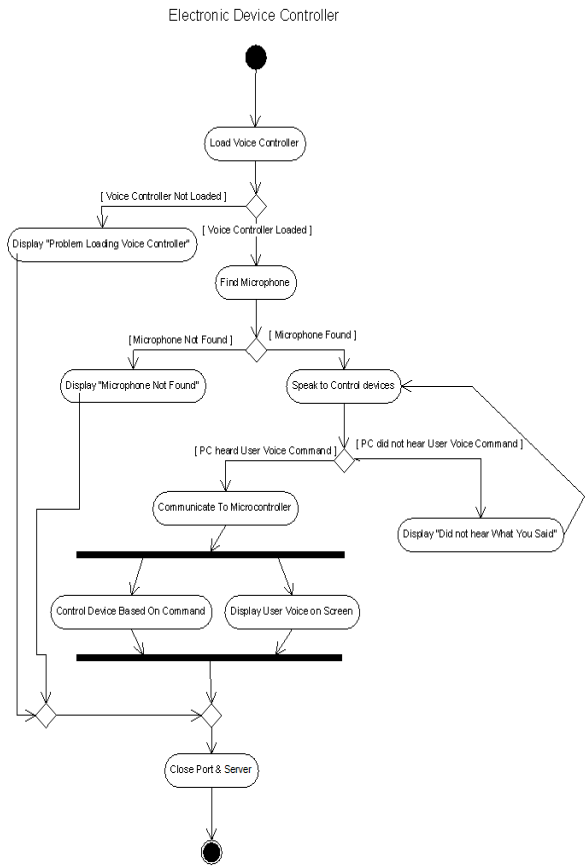

Figure 3. Electronic device controller

*a)  Electronic device control using Voice command*

Unlike other common systems, this subsystem is open source & development cost is nothing comparing to others. The voice reorganization is not bound to any specific voice command. The system does not need any training to identify user command. To apply this action in voice recognition, we used open source voice reorganization called sphinx that made voice reorganization easy. But some problem is non-avoidable like noise or slow in recognizing.

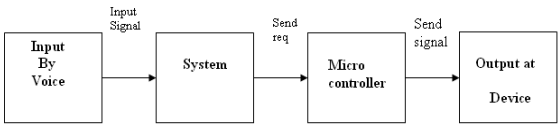

Figure 4. Block Diagram of Electronic device control by voice.

*b)  Electronic device control using Bluetooth*

This subsystem is mainly responsible for controlling electrical devices via Bluetooth. We firstly tried to implement Bluetooth stack in desktop using Java SE but then we found there is no Bluetooth support in JAVA SE. Finally we used Native API to overcome this problem. This makes it non portable means not platform independent.

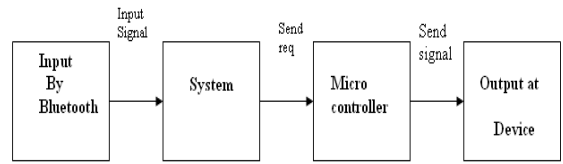

Figure 5. Block Diagram of Electronic device control using Bluetooth.

*c)  Electronic device control using Website*

This subsystem enables user to control his/her home electronic devices from remote area without installing any software on client side. The main challenge here was using comm API (serial port communication) from web page. Since for security reason, java does not allow direct access to Comm PORT. To overcome this challenge we used **Servlet** and **JSP** and **Java SE** class with comm Port implementation that communicates with microcontroller via web page.

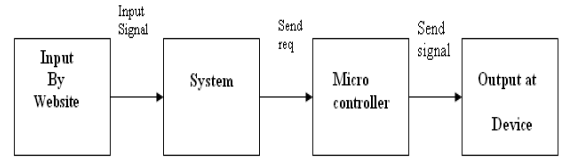

Figure 6. Block Diagram of Electronic device control using Web site.

*2)  IR Security System*

IR security system is a virtual security guard where the IR device throws the infrared ray in a particular area (i.e. secured zone). If any unauthorized person enters in that area which was totally unexpected to the Home Owner, IR security system informs to the user (i.e. Home owner) by sending SMS that "someone entered in that specific area/secured zone". Moreover system will also capture & stores the picture of that unauthorized person for further actions. Furthermore the system will also generate an alarm in the room informing "someone in the room". So that instant action can be taken. There may raise a question what if an authorized person enters to the room while system is running? Well the solution is the user can stop the service for a certain period of time by pressing stop scanning from system server or anywhere via internet. So the system can be used as a virtual security guard that ensures the security of the valuable things & save from harm.

Pseudo code for IR Security System is given below:

```
IRSecuritySystem

1       Start Server PC
2       Start System Software
3       Give User Mobile number //for SMS
4       If (number is given)
5           Store Mobile number to PC for SMS
6           Get scan command
7           If(start scanning command is given)
8               Start scanning for unauthorised person
9               If(unauthorized person detected)
10                  Send warning message to user via SMS
11                  Store Image in PC for future identification
12                  Speaker alarm "Someone in the room"
13                  If(System windows not closed)
14                      Go to line 8
15                  Else
16                      System closed
17                      Go to line 28
18              Else
19                  Continue scanning for unauthorized person
20                  Go to line 9
21          Else if(Stop command is given)
22              Stop scanning
23              Go to line 6
24          End if
25      Else
26          Wait until number is given
27          Go to line 3
28      End IRSecuritySystem
```

The activity diagram of IR security system has given in figure 7.

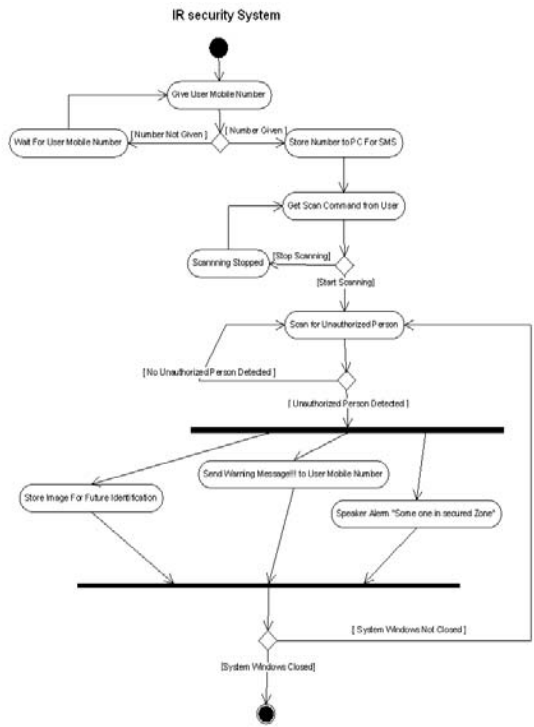

Figure 7. Activity diagram of IR security system

*3) Remote Video Surveillance System/ Remote eye*

Remote video surveillance is a system by which user can monitor a particular selected region from remote distance. A camera is placed on that surveillance region. User will login to the system from remote area from a Mobile via GPRS or internet by putting Server IP address & then give command for receiving continuous video stream of that specific region (i.e. home). Camera will capture the picture of that particular zone & send the continuous bit stream to the user. User will be able to see the live video of his/her home. The system will not let user feel alone while s/he is outdoor or abroad. User will feel as if he is inside the home. User can login & monitor his/her home whenever s/he wants (i.e. 24/7). So the user doesn't need to be tensed about his/her home. The Remote Video Surveillance System allows user to be available in home virtually. The system works as remote eye of home owner.

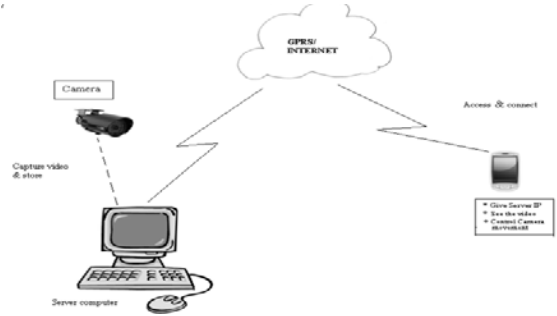

Figure 8. Overview of remote video surveillance

The activity diagram of remote video surveillance is given in figure 9.

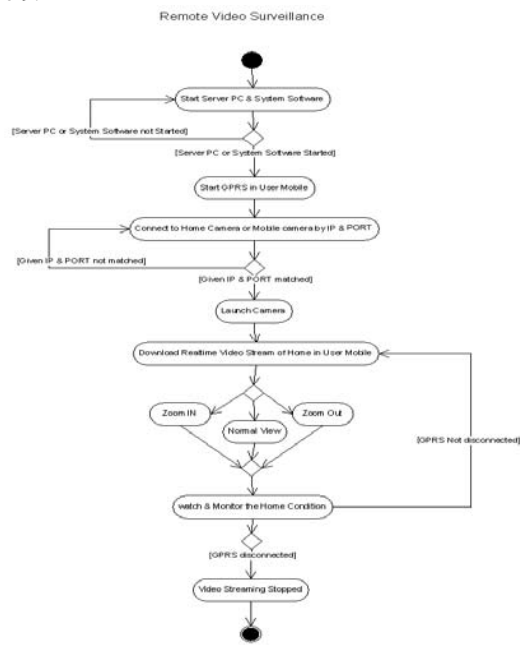

Figure 9. Activity diagram of remote video surveillance.

*4) Web Desktop*

Web desktop is internet based software system by which user can control his/her computer remotely, from anywhere in the world with the help of web browser. The system is mainly built to control the **DHS's** main server from remote place. So the user can control any subsystems (i.e. activate/deactivate IR security or virtual mobile system etc) . Using this software any internet enabled computer will act as a remote controller & communicates with the PC (via the Internet). User can do many tasks such as open and close audio/video file, software, and system software like task manager and even shut down, restart or log off the computer from a remote place. Web Desktop also gives the facility to view user's computer contents, write document, DOS command execution, quick launch facility of regular used applications. Moreover all kind of mouse event work can be done by this software. Now you don't need to carry your data or laptop. All you just need is to remember your IP address & one internet enabled computer in anywhere in the world. Pseudo code for Web desktop is given below.

```
WebDesktopActivity
1       Start Server PC
2       Start Client PC
3       While (Client Pc & Server PC or system software not closed) do
4            open internet Browser on Client PC
5            Give Server PC IP on address bar
6            While (internet Browser Not Closed)
7            For each (IP in Possible IPs) do
8                 If (Server PC IP is equal to Given IP) then
9                     Connect to the server computer
10                    Load & Display Server computer Desktop on browser
11                    if (any desktop icon Clicked on browser) then
12                        Calculate the X, Y position of that Icon in server PC
13                        Click Server PC Icon after actual measure
14                        Complete the Tasks
15                        Go to line 6
16                    Else
17                        Wait for user to Click
18                        Go to line 11
19                Else
20                    Go to line 5
21            End For
22            End While
23      End While
24      End WebDesktopActivity
```

The activity diagram of web desktop is given in figure 10.

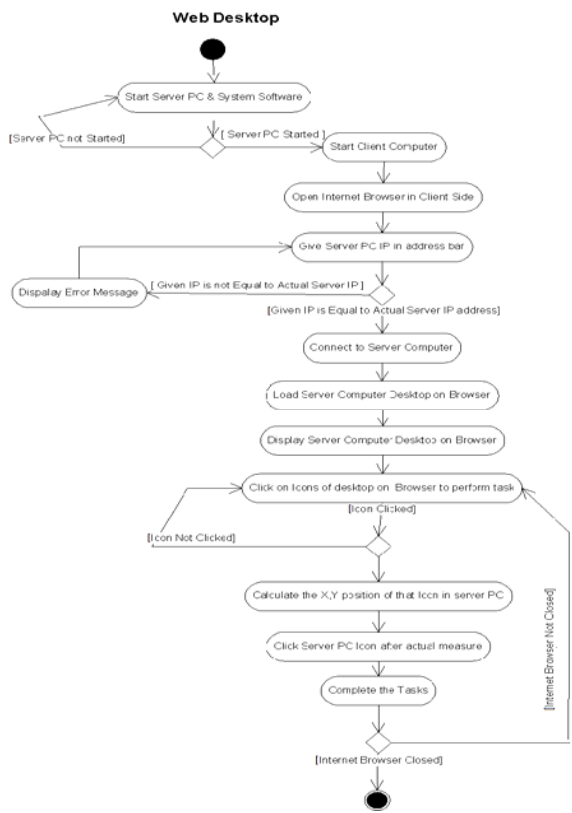

Figure 10. activity diagram of web desktop

### 5) *Virtual Mobile*

Virtual mobile application is a Universal mobile Driver by which a user can exactly perform some same task(e.g. Remote call, Phone book access, SMS read-write ) of a original mobile device from his/her PC's virtual mobile or another mobile. The virtual mobile send a request to the original mobile device via Bluetooth. Once the request is accepted by the original mobile, then connection is established. After that the requester mobile can perform exactly the same task like original one. Finally after the task completion or during the task operation if original mobile owner wants to disconnect the connection between virtual & original mobile she can do it by switching off the Bluetooth.

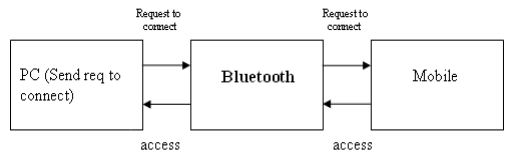

Figure 11. Block Diagram for Virtual Mobile.

The activity diagram of virtual mobile is given in figure 12.

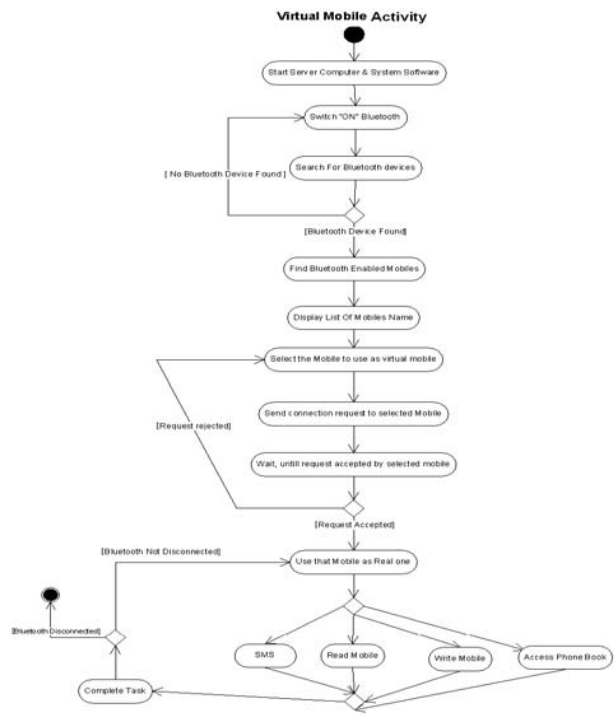

Figure 12. Activity diagram of virtual mobile

## 4. SIMULATED OUTPUT OF OUR PROPOSED SYSTEM

We have designed and developed all of our proposed models. Also we have simulated our proposed model. Some snapshots of our simulation have given below.

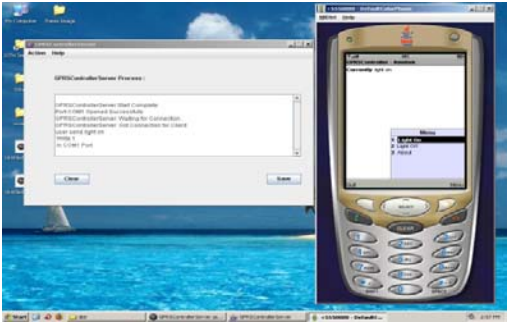

Figure13. Light On and off from mobile phone

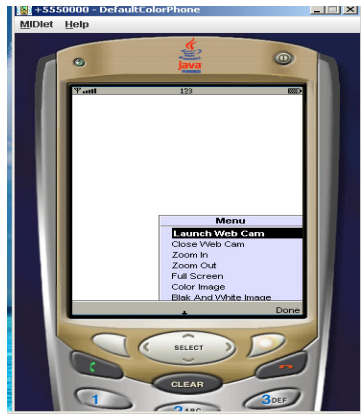

Figure14. Controlling webcam from mobile phone.

## 5. BENIFITS

Our digital home system provides a low cost but luxury automated home system. It has some very significance benefits.

- It makes our home automated in terms of manual system that makes life simpler & also add some luxury features.
- The Digital Home System helps the user to lead a luxurious life. User can now control Light or fan AC or any electronic devices at home by voice command, Bluetooth, GPRS, and SMS.
- Time is the most important factor of our life. This system reduced our time consumption as the manual process takes more time than automated one.
- The system is giving the portability to the users. User is now able to access PC being anywhere in the world.
- System is doing most of the regular tasks (e.g. light/fan on/off) of us (i.e. human being). So now there is more time to relax
- Unlike the manual process, user does not need to be at home physically. All s/he needs to do is just use his/her mobile phone to get the job done.
- Security is a big issue in our daily life. The system ensures some security features us by informing up-to date information to us.
- System detects intruder's entry at home, after that automatically informing us by sending SMS & store the image in server Computer. So that we can take action immediately & identify the intruder later. So the home security increased.
- Now we don't need to be worry or tensed by thinking about home situation. We can keep an eye on our home 24/7 from anywhere in the world.
- All the electronic devices are controlled by the system. So the risk of electric shock is totally reduced. We can now live a tension free life.
- We have multiple IR in our IR security system. So that no single object can enter in secured zone.
- We have developed the mobile based system. User will be able to access & operate DHS from their mobile device.
- Our system will open/close door automatically.
- Light will be automatically on/off based on persons in out in the room.
- The system will be able to control water level of water tank.
- Public IP preference will be optional in web desktop and remote surveillance system
- We added the GPS feature to identify the exact location of precious metals(i.e. gold, Diamond, jewelries etc )
- The system is developed by using very low cost PIC microcontroller which reduced the development cost enormously.
- Virtual mobile will be accessible by GPRS/INTERNET instead of only Bluetooth.
- Both mobile and PC can be used.
- Very low cost digital home system.
- Our system will open/close door automatically.
- Light will be automatically on/off based on persons in out in the room.

## 6. CONCLUSION

The "Controlling and Securing a Digital Home, using Multiple Sensor Based Perception System Integrated with Mobile and Voice Technology" invention provides better security & luxury features to home automation at a reasonable cost and accepted by global systems to control remotely from any computer in the world. As we have included mobile technology and voice technology in our system, this will be more convenient and reliable for the user.